\documentclass[aps,prd,amsfonts,amsmath,nofootinbib,preprint,tightenlines,showpacs]{revtex4-1}

\def\be{\begin{equation}}
\def\ee{\end{equation}}
\def\bea{\begin{eqnarray}}
\def\eea{\end{eqnarray}}
\def\bml{\begin{subequations}}
\def\blea{\bml\bea}
\def\elea{\end{eqnarray}\end{subequations}}

\def\intoi{\int_0^\infty}
\def\intou{\int_0^u}
\def\intov{\int_0^v}
\def\Eone{\text{E}_1}
\def\Ei{\mathop{\text{Ei}}}
\def\Er{E_R}
\def\El{E_L}
\def\Eext{E_{\text{ext}}}
\def\Eint{E_{\text{int}}}
\def\Jrt{J^u_{\text{tot}}}
\def\Jlt{J^v_{\text{tot}}}
\def\Jr{J^u}
\def\Jl{J^v}
\def\Jsr{J^u_R}
\def\Jsl{J^v_L}

\def\Jintr{J^u_{\text{int}}}
\def\Jintl{J^v_{\text{int}}}
\def\Irs{I^u_{\text{strip}}}

\def\Real{\mathop{\text{Re}}}
\pdfoutput=1

\usepackage{graphicx}
\graphicspath{ {./Figures/} }

\begin{document}

\title{Electromagnetic back-reaction from currents on a straight string}

\author{Jeremy M. Wachter}
\email{Jeremy.Wachter@tufts.edu}
\author{Ken D. Olum}
\email{kdo@cosmos.phy.tufts.edu}
\affiliation{Institute of Cosmology, Department of Physics and Astronomy,\\ 
Tufts University, Medford, MA 02155, USA}

\begin{abstract}
Charge carriers moving at the speed of light along a straight,
superconducting cosmic string carry with them a logarithmically
divergent slab of electromagnetic field energy.  Thus no finite local
input can induce a current that travels unimpeded to infinity.
Rather, electromagnetic back-reaction must damp this current
asymptotically to nothing.  We compute this back-reaction and find
that the electromagnetic fields and currents decline exactly as
rapidly as necessary to prevent a divergence.  We briefly discuss the
corresponding gravitational situation.
\end{abstract}


\maketitle

\section{Introduction}\label{sec:intro}

Cosmic strings are effectively-one-dimensional objects that may have
formed as topological defects during spontaneous symmetry breaking in
the early universe, or as string theory objects at the end of ``brane
inflation'' \cite{Sarangi:2002yt}.  For a review, see
Ref.~\cite{CSOTD}.  Strings may be of cosmological and astronomical
significance in a number of ways, such as producing CMB anisotropies
\cite{cmbanis1,cmbanis2}, sourcing gravitational waves \cite{gravrad},
or gravitationally lensing astrophysical objects \cite{microlens}.
The strongest constraints on cosmic strings come from non-observation
of gravitational waves in pulsar timing experiments
\cite{Sanidas:2012ee,gmulim}.

Superconductivity is a common property of cosmic strings. In many
models, strings may carry a conserved current comprised of massless
charge carriers \cite{supercon}.  In general the charge carriers might
carry some charge other than the usual electromagnetic one, and so
might be coupled to some other field or no field at all, but here we
will consider only the case where the charge carriers on the string are
electromagnetically charged.  Superconductivity gives additional
possibilities for detection of cosmic strings, such as emission of
charge carriers that decay into cosmic rays \cite{Berezinsky:2009xf}
and emission of bursts of electromagnetic radiation
\cite{Vilenkin:1986zz}.

In this paper we will consider the electromagnetic effects of charges
and currents on a straight, static string.  We will initially treat
the string as a line of zero width, but later we will introduce, in a
heuristic way, a length $\delta$ characteristic of the width of the
string core.

In the absence of external fields, there is an exact solution giving
the fields of charges and currents on a string \cite{Aryal:1987pw}.
For a string lying on the $x$ axis, in cylindrical coordinates,
\bml\label{eqn:stat}\bea
\mathbf{E}(x,\rho,\theta)&=&2J^t(x)\rho^{-1}\hat\rho\,,\\ \mathbf{B}(x,\rho,\theta)&=&2J^x(x)\rho^{-1}\hat\theta\,,
\elea
where $J^x$ is the electric current flowing in the positive $x$
direction and $J^t$ the electric charge.  Since there is no electric
field pointing in the $x$ direction, there is no back-reaction and the
currents do not dissipate \cite{Aryal:1987pw}.  Note, however, that the
fields in Eq.~(\ref{eqn:stat}) have divergent total energy.  Since the
fields go as $1/\rho$, the energy density $\varepsilon$ goes as
$1/\rho^2$. The energy per unit length $\mathcal{E}$ is then
\be
\mathcal{E}=\int\varepsilon\rho\, d\rho\, d\theta
\propto\int\frac{1}{\rho} d\rho\,,
 \ee 
which diverges logarithmically as $\rho\to\infty$, even for a
finite-width string.

\begin{figure}
\centerline{\includegraphics[scale=0.3]{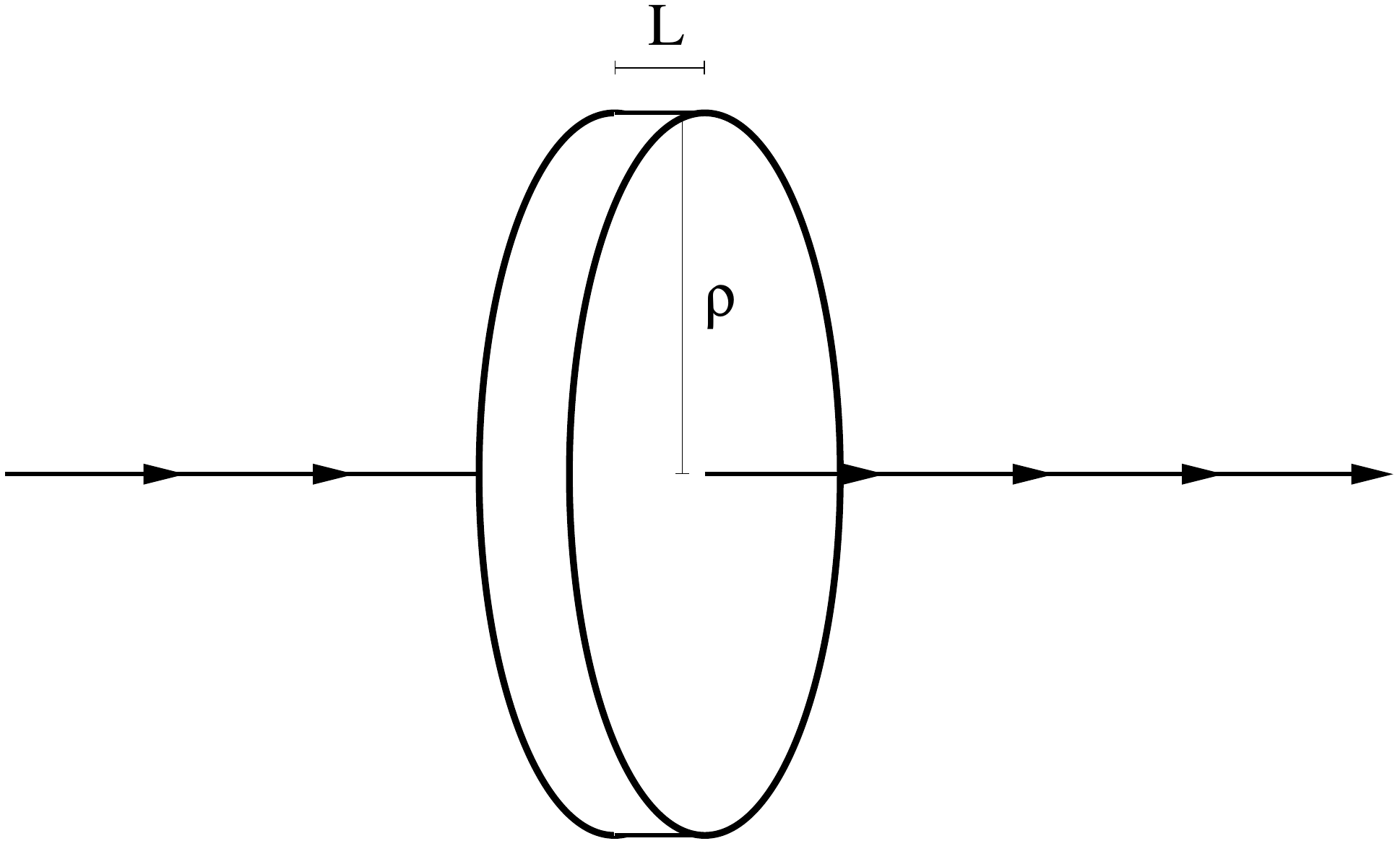}}
\caption{A train of charge carriers of length $L$, moving at the speed
  of light, carries with it a disk-shaped slab of electromagnetic
  energy.}\label{fig:slab}
\end{figure}

Now suppose we attempt to induce a current on a string, by applying
some external electric field for some finite time in a localized
region.  If the field points toward the right, we will produce some
positive charge carriers moving to the right and some negative charge
carriers moving to the left.  In the absence of back-reaction, these
charge carriers would move unimpeded to infinity, as shown in
Fig.~\ref{fig:slab},
and the solution would approach the exact solution of
Eq.~(\ref{eqn:stat}).  But that solution is impossible to obtain,
because our procedure can only supply a finite energy to the string,
leaving us with a paradox.

Clearly back-reaction must decrease the currents.  In fact, it must
decrease the current and charge density asymptotically to zero,
because the final solution to this problem cannot allow for any
stationary currents in the final state.  If any part of the initial
train of charge carriers continues to propagate unchanged along the
string (at some constant, nonzero population density), then the
paradox still exists.

Indeed, we will find that the initial current, or more accurately,
the induction of the initial current, leads at later times to an
electric field that points oppositely to the applied field, which
reduces the current.  But how can the current be reduced?  As the
charge carriers are massless, they move at the speed of light and
cannot be slowed, and by charge conservation they cannot be destroyed.
Instead, the effect of the field is to remove charge carriers from the
initial current and create charge carriers with the same charge but
the opposite direction of motion.  Alternatively, we may say that the
charge carriers scatter off the field and reverse their direction of
motion.  We will find that the electric fields and currents produced
by back-reaction drop off exactly as quickly as is necessary to
prevent this paradox, and that they eventually go to zero at infinity.

The rest of the paper is organized as follows: In
Sec.~\ref{sec:inteq}, we find an integral equation for the electric
fields and currents resulting from a $\delta$-function source. In
Sec.~\ref{sec:fcsgl}, we solve this equation for the electric fields
and currents on the future light cone of the point where the source is
applied. In Sec.~\ref{sec:fcint}, we find the electric field and
current in the interior of the future light-cone. In
Sec.~\ref{sec:stripcur}, we find the current contained in some finite
strip of the spacetime diagram. In Sec.~\ref{sec:totcur}, we examine
the effects of a general source. We conclude in Sec.~\ref{sec:concl}
with a discussion of our results and how they might be relevant to
gravitational back-reaction on a string subject to some spatial
displacement.

We use metric signature $(+,-,-,-)$ and work in units where $c=1$ and
$\hbar=1$.  We express electromagnetic quantities in Gaussian units,
so the squared electric charge is $q^2=\alpha\approx1/137$.

\section{The induced electric field and the resulting current}\label{sec:inteq}

Charges and currents on a string give rise to electric fields, which
can induce further currents.  Consider a string lying along the $x$
axis with some linear charge density $J^t(x,t)$ and current
$J^x(x,t)$.  The 3-dimensional charge-current density 4-vector is then
$j^\mu(X)=J^\mu(x)\delta(y)\delta(z)$, and the induced electromagnetic
potential in Lorenz gauge is
\be\label{eqn:fpg}
A^\mu(X)=\int G(X,X')j^\mu(X') d^4X'
= \int G(x,t,x',t')J^\mu(x',t')dx' dt'\,,
\ee
where $G$ is the retarded Green's function.

We will henceforth be interested only in the electric field at
locations on the string, and only in the component in the $x$
direction, because this is the field that leads to changes in the string
currents.  It is given in terms of the potential by
\be\label{eqn:eft}
E_x(x,t)=F^{10}=-\partial_xA^t-\partial_tA^x\,.
\ee
After this, we will drop the subscript $x$ on E.

Applying the derivatives of Eq.~(\ref{eqn:eft}) to Eq.~(\ref{eqn:fpg}) gives
\be\label{eqn:xtiraw}
E(x,t)=-\int\left(J^t(x',t')\partial_x G(x,t,x',t')+J^x(x',t')\partial_t G(x,t,x',t')\right)
dx' dt'\,.
\ee
Since the Green's function depends only on $x-x'$ and $t-t'$, we can write $\partial_x G=-\partial_{x'} G$, $\partial_t G=-\partial_{t'}G$.  Then we integrate by parts to find
\be\label{eqn:egj}
E(x,t)=-\int G(x,t,x',t')\left(\partial_{x'} J^t(x',t') + \partial_{t'} J^x(x',t')\right)dx' dt'\,.
\ee

The effect of an electric field on a superconducting string is to
induce a current \cite{supercon,CSOTD},
\be\label{eqn:cee}
\partial_xJ^t+\partial_tJ^x=q^2E_x\,.
\ee
Putting Eq.~(\ref{eqn:cee}) in Eq.~(\ref{eqn:egj}), we see that an electric field applied to the string at time $t'$ leads to an additional electric field at any later time $t$,
\be\label{eqn:ee1}
E(x,t)=-q^2\int G(x,t,x',t')E(x',t')dx'dt'\,.
\ee
Equation~(\ref{eqn:ee1}) is incomplete, because on the right-hand side $E$ means the total electric field, including external sources, whereas on the left it means only the field induced by the string current.  If we include some external field $\Eext$, we have
\be\label{eqn:ee2}
E(x,t)=\Eext-q^2\int G(x,t,x',t')E(x',t')dx'dt'\,.
\ee
This integral equation allows us, in principle, to find the electric field in the presence of a string responding to any given applied external field.

In the absence of any electric field, charge carriers move freely at the speed of light in the positive and negative $x$ directions on the string.  It is therefore convenient to change to a null coordinate system with
\blea
u&=&\frac{t+x}{\sqrt{2}}\,,\\
v&=&\frac{t-x}{\sqrt{2}}\,.
\elea
The retarded Green's function in two dimensions in these coordinates is
\be
G=\frac{\delta(v-v')}{u-u'}+\frac{\delta(u-u')}{v-v'}\,,
\ee
and Eq.~(\ref{eqn:ee2}) becomes
\be\label{eqn:ee3}
E(u,v)=\Eext-Q\left(\intou\frac{E(u',v)}{u-u'}du'
+\intov\frac{E(u,v')}{v-v'} dv'\right)
\ee
where $Q=q^2$. We interpret this integral equation as follows: The electric field at each point is given by an integral over all electric fields on that point's past light cone. The integral over $u'$ gives the contribution from all fields earlier in time and to the left, while the integral over $v'$ gives the contribution from all fields earlier in time and to the right.

Unfortunately, the integrals in Eq.~(\ref{eqn:ee3}) have a divergence
at the upper limit of integration, which results from our treatment of
the string as a one-dimensional object, giving it infinite
self-inductance.  To fix this problem, we add in a constant $\delta$,
characteristic of the string's width, to both of the denominators,
which gives
\be\label{eqn:uvint}
E(u,v)=\Eext-Q\left(\intou\frac{E(u',v)}{u-u'+\delta}
du'+\intov\frac{E(u,v')}{v-v'+\delta}dv'\right)\,.
\ee

We now split up the current into parts that propagate to the right and to the left,
\blea
\Jr&=&\frac{J^t+J^x}{\sqrt{2}}\,,\\
\Jl&=&\frac{J^t-J^x}{\sqrt{2}}\,.
\elea
Using the continuity equation, $\partial_tJ^t+\partial_xJ^x=0$, and
Eq.~(\ref{eqn:cee}), we find
\be
\partial_u\Jr=-\partial_v\Jl=\frac{Q}{2}E\,.
\ee
A positive value of $\Jr$ represents positive charge carriers moving to the right.  In the absence of any electric field, these carriers move freely at the speed of light, so $\Jr$ does not depend on $u$. A positive value of $\Jl$ represents positive charge carriers moving to the left. In the absence of any electric field, these carriers move freely to the left, so $\Jl$ does not depend on $v$.

Once we have determined the electric field, we can determine the current via
\bml\label{eqn:Je}\bea
\Jr&=&\frac{Q}{2}\intou E(u',v)du'\,,\\
\Jl&=&-\frac{Q}{2}\intov E(u,v')dv'\,.
\elea

\section{The electric field and current on the null lines}\label{sec:fcsgl}

We will start by considering the case where the applied field is a $\delta$-function at the origin, $\Eext = \delta(x)\delta(t) =\delta(u)\delta(v)$.  In this case, there will be singular currents and fields on the $u$ and $v$ axes and nonsingular currents in the chronological future of the origin, $u,v > 0$.  We will find the singular currents and fields first.

We write the total field,
\be\label{eqn:Eparts}
E=\Eext+\Eint+\Er+\El\,,
\ee
where $\Eint$ is the ``interior'' electric field in the chronological future of the origin and $\Er,\,\El$ are the singular fields on the positive $u$ and $v$ axes, respectively. A positive value of $\Er$ or $\El$ represents an electric field pointing rightward and $\El(u,v)=\Er(v,u)$ by symmetry.

To find $\Er$, we use Eq.~(\ref{eqn:Eparts}) in Eq.~(\ref{eqn:uvint})
and only consider points $(u,v)$ on the positive $u$-axis. Since there
is nothing in the past to the right of the the right null line, the
second term of Eq.~(\ref{eqn:uvint}) does not contribute. Furthermore,
we see that $\Eint$ and $\El$ cannot be in the past of the $u$-axis.
Thus we use $E = \Eext + \Er$ on both sides of Eq.~(\ref{eqn:uvint}),
and integrate the resulting $\delta(u)$, to get
\be\label{eqn:Er1}
\Er(u,v)=-Q\left(\frac{\delta(v)}{u+\delta}+\intou\frac{\Er(u',v)}{u-u'+\delta}du'\right)\,.
\ee
We can separate out the $\delta$-function in $\Er$ by writing
\be\label{eqn:Ef}
\Er(u,v)=-f(u)\delta(v)\,.
\ee
Then $f$ satisfies
\be
f(u)=h(u)-(f*h)(u)\label{eqn:fint}\,.
\ee
where
\be\label{eqn:hdef}
h(z)=\frac{Q}{z+\delta}\,,
\ee
and $*$ indicates a convolution, defined by
\be
(f*h)(u)=\intou f(u')h(u-u')du'\,.
\ee
Then, with the Laplace transforms
\bml\label{eqn:H}\bea
\mathfrak{L}[f](w)&=&F(w)\,,\\
\mathfrak{L}[h](w)&=&H(w)=Qe^{\delta w}\Eone(\delta w)\,,
\elea
where $\Eone$ is the exponential integral function, we perform the transform of Eq.~(\ref{eqn:fint}) to get
\be
F(w)=H(w)-F(w)H(w)\,,
\ee
and so
\be\label{eqn:F}
F(w)=\frac{H(w)}{1+H(w)}=\frac{Qe^{\delta w}\Eone(\delta w)}{1+Qe^{\delta w}\Eone(\delta w)}\,.
\ee
We now take the inverse Laplace transform of $F(w)$.  As discussed in
the appendix, $F(w)$ has a branch cut on the negative real axis and
no singularities in the right half-plane, and does not diverge at
$w=0$.  Thus the inverse Laplace transform is
\be
f(u)=\frac{1}{2\pi i}\int_{-i\infty}^{+i\infty}F(w)e^{wu}dw\,.
\ee
Furthermore, the integrand falls quickly enough at infinity that we
can deform the contour of integration to enclose only the negative
real axis. Thus,
\be\label{eqn:deform}
f(u)=\frac{1}{2\pi i}\intoi
(F(-a-i0)-F(-a+i0))e^{-au}da\,.
\ee
Now using the property of $\Eone$ that
\be
\Eone(-x\pm i0)=-\Ei(x)\mp i\pi\,,
\ee
and changing variables to $\alpha=\delta a$, we get
\be\label{eqn:fint1}
f(u)=\frac{Q}{\delta}\intoi\frac{e^{-\alpha(u/\delta+1)}}{(1-Qe^{
-\alpha}\Ei(\alpha))^2+(Q\pi e^{-\alpha})^2}d\alpha\,.
\ee

Equations~(\ref{eqn:Ef},\ref{eqn:fint1}) give an exact solution to Eq.~(\ref{eqn:Er1}), but the integral cannot be done in closed form, so we must make approximations. Since we inserted $\delta$ by hand we do not expect this expression to be good for $u$ comparable to $\delta$, so we take $u\gg\delta$.  Then the exponential term in the numerator suppresses the integrand unless $\alpha\ll 1$, which means we can neglect factors of $e^{-\alpha}$ everywhere, and use the small-argument approximation $\Ei(\alpha)=\ln \alpha+\gamma$, where $\gamma$ is the Euler-Mascheroni constant. This gives us a simplified integral equation,
\be\label{eqn:fint2}
f(u)=-\frac{Q}{\delta}\intoi\frac{e^{-\alpha
    u/\delta}}{(1-Q(\ln \alpha+\gamma))^2+(Q\pi)^2}d\alpha\,.
\ee

Now we make the much stronger approximation that $\ln(u/\delta)\gg 1$,
meaning that $u$ is many orders of magnitude larger than
$\delta$. This is justified because we are concerned with a paradox
that occurs only when logarithms become large.  Then we observe that
for the great majority of the range of integration before the integral
is cut off by the exponential at $\alpha\sim \delta/u$, $\ln\alpha$
has a value near $\ln(\delta/u)$.  Thus we approximate the denominator
in Eq.~(\ref{eqn:fint2}) by its value at $\alpha=\delta/u$ and perform
the integral.  We can ignore $\gamma$ by comparison with
$\ln(\delta/u)$, and the second term in the denominator by comparison
with the first, to get
\be\label{eqn:f}
f(u)=-\frac{Q}{u}\frac{1}{(1+Q\ln(u/\delta))^2}\,,
\ee
which gives, by Eq.~(\ref{eqn:Ef}),
\be\label{eqn:Erfinal}
\Er(u,v)=\frac{Q}{u}\frac{\delta(v)}{(1+Q\ln(u/\delta))^2}\,.
\ee
Note that we do not consider $Q\ln(\alpha)$ to be much greater than 1, because $Q$ may be small.

We are now prepared to find the right-moving singular current. From Eqs.~(\ref{eqn:Je},\ref{eqn:Ef}), we get
\be\label{eqn:Je1}
\Jsr(u,v)=\frac{Q}{2}\intou \left(\Eext(u',v)+\Er(u',v)\right)du'
=-\frac{Q}{2}\delta(v)\intou \left(\delta(u')-f(u')\right) du'\,.
\ee

Because our approximations are not accurate for small $u$, it is better to change the range of integration, as follows.  By the definition of the Laplace transform, $\intoi f(u) du = F(0) = 1$, so $\intoi (\delta(u)-f(u)) du = 0$, and thus
\be\label{eqn:Je2}
\Jsr(u,v)=\frac{Q}{2}\delta(v)\int_u^\infty f(u') du'\,.
\ee
This means that the current falls asymptotically to zero at late times, which is what we expect from our general arguments above.

This integral is easily done, and tells us that the right-going singular current is
\be\label{eqn:Jsr}
\Jsr(u,v)=\frac{Q}{2}\frac{\delta(v)}{1+Q\ln(u/\delta)}\,.
\ee

The expressions for the left-going singular electric field and current are found from $\El(u,v)=\Er(v,u)$ and $\Jsl(u,v)=-\Jsr(v,u)$ and Eqs.~(\ref{eqn:Erfinal},\ref{eqn:Jsr}). We see that the singular currents are always right-going and go to zero as $u$ or $v$ goes to infinity. The singular electric fields, after the initial kick, are negative and going to zero faster than the singular currents. 
Examining the singular current more closely along either the $u$ or $v$ axis, we see that half of the initial current has been scattered away when the varying coordinate reaches $\delta e^{1/Q}$. We will discuss the scale of this decline in Sec.~\ref{sec:concl}.

This solution for the singular currents solves the problem of the
divergence of the energy contained in the electric field. The currents
decline exactly as quickly as is required to cancel the logarithmic
divergence along the null lines.

\section{The electric field and current on the interior}\label{sec:fcint}

We are now interested in finding the electric field and currents in
the chronological future of the source, $\Eint$, $\Jintr$, and
$\Jintl$. We begin to solve for $\Eint$ via the same process as for
the singular electric field.  We note that $\Eext$ is not on the past
light-cone of any point in the interior and that we need both
integrals from Eq.~(\ref{eqn:uvint}), which we now write as
\be
\Eint(u,v)=f(u)h(v)+f(v)h(u)-(\Eint(\cdot,v)*h)(u)-(\Eint(u,\cdot)*h)(v)\,.
\ee
Where we solved for the singular field using one Laplace transform
from the domain $u\to w$, here we will solve by taking a double
Laplace transform from $u\to w$ and $v\to y$.  We let
$\mathcal{H}=\mathfrak{L}[\Eint]$, so
\be
\mathcal{H}(w,y)=F(w)H(y)+F(y)H(w)-\mathcal{H}(w,y)(H(w)+H(y))
\ee
and
\be\label{eqn:HH}
\mathcal{H}(w,y)=\frac{F(w)H(y)+F(y)H(w)}{1+H(w)+H(y)}
= \frac{1}{1+H(w)+H(y)}-\frac{1}{1+H(w)}-\frac{1}{1+H(y)}+1\,.
\ee

The properties of $\mathcal{H}$ are discussed in the appendix.  There
are no singularities in the right half-plane, so the double inverse
Laplace transform is
\be\label{eqn:L2}
\Eint(u,v)=-\frac{1}{4\pi^2}\int_{-i\infty}^{+i\infty}e^{wu}dw
\int_{-i\infty}^{i\infty}e^{yv}dy\mathcal{H}(w,y)
\ee
and we can deform the contours as before to find
\bea\label{eqn:deform2}
\Eint(u,v)=-\frac{1}{2\pi^2}\intoi\intoi[&&
\mathcal{H}(-a-i0,-b-i0)-\mathcal{H}(-a+i0,-b-i0)\\
&&-\mathcal{H}(-a-i0,-b+i0)+\mathcal{H}(-a+i0,-b+i0)]e^{-au-bv}dadb\,.\nonumber
\eea
Note that only the first term of $\mathcal{H}$ depends on both $w$ and $y$; as a result, the subtractions in Eq.~(\ref{eqn:deform2}) cancel the other terms in pairs. We change variables to $\alpha=\delta a$ and $\beta=\delta b$, then simplify to obtain
\be
\Eint(u,v)=\frac{2}{\delta^2}\intoi\intoi\frac{CQ^2e^{-\alpha(1+u/\delta)-\beta(1+v/\delta)}}{C^4+2C^2Q^2\pi^2\left(e^{-2\alpha}+e^{-2\beta}\right)+Q^4\pi^4\left(e^{-2\alpha}-e^{-2\beta}\right)^2}d\alpha d\beta\,,
\ee
where $C=1-Q\left(e^{-\alpha}\Ei(\alpha)+e^{-\beta}\Ei(\beta)\right)$. We now make the approximations $u,v\gg\delta$, which give
\be
\frac{2Q^2}{\delta^2}\intoi\intoi\frac{e^{-(\alpha u+\beta v)/\delta}}{(1-Q(\ln(\alpha\beta)+2\gamma))^3+4Q^2\pi^2(1-Q(\ln(\alpha\beta)+2\gamma))}d\alpha d\beta\,.
\ee

Now, we apply the strong approximation $\ln(u/\delta),\ln(v/\delta)\gg 1$, as in Sec.~\ref{sec:fcsgl}. Furthermore, the cubic term is much greater than the linear term, and so we ignore the linear term in its entirety. Thus, the interior electric field is
\be
\Eint(u,v)=\frac{2Q^2}{uv}\frac{1}{(1+Q\ln(uv/\delta^2))^3}\,.
\ee
We see that the interior electric field always points to the right,
going to zero when $u$ or $v$ goes to infinity. It declines faster
than the singular electric field.  

We may find the interior currents using Eq.~(\ref{eqn:Je}).  In the
solution for $\Jintr$, we see immediately that $\Er$ and $\Eext$ will
not contribute, because we are interested in points for which $v>0$.
The integral of $\El(u',v)$ is trivial and yields $-f(v)$. From the
definition of the Laplace transform,
\be
\intoi\Eint(u,v)du=\frac{1}{2\pi i}\int_{i\infty}^{i\infty}e^{yv}dy\frac{F(0)H(y)+F(y)H(0)}{1+H(0)+H(y)}\,.
\ee
Then, because $H(0)=\infty$, this simplifies to
\be\label{eqn:Eintzero}
\intoi\Eint(u,v)du=\frac{1}{2\pi i}\int_{i\infty}^{i\infty}F(y)e^{yv}dy=\mathfrak{L}^{-1}[F](v)=f(v)\,.
\ee
We see that
\be
\intoi E(u',v)du'=0\,,
\ee
and therefore
\be
\intou E(u',v)du'=-\int_u^{\infty}E(u',v)du'\,.
\ee
Using this in Eq.~(\ref{eqn:Je}), we find
\be
\Jintr(u,v)=\frac{Q}{2}\intou \Eint(u',v)du'=-\frac{Q^2}{2v}\frac{1}{(1+Q\ln(uv/\delta^2))^2}\,,
\ee
and by symmetry
\be
\Jintl(u,v)=\frac{Q^2}{2u}\frac{1}{(1+Q\ln(uv/\delta^2))^2}\,.
\ee
We see that the interior currents are always left-going and go to zero as either $u$ or $v$ goes to infinity. They decline faster than the singular currents, and so the paradox is also solved for the interior.

\begin{figure}
\centerline{\includegraphics[scale=0.75]{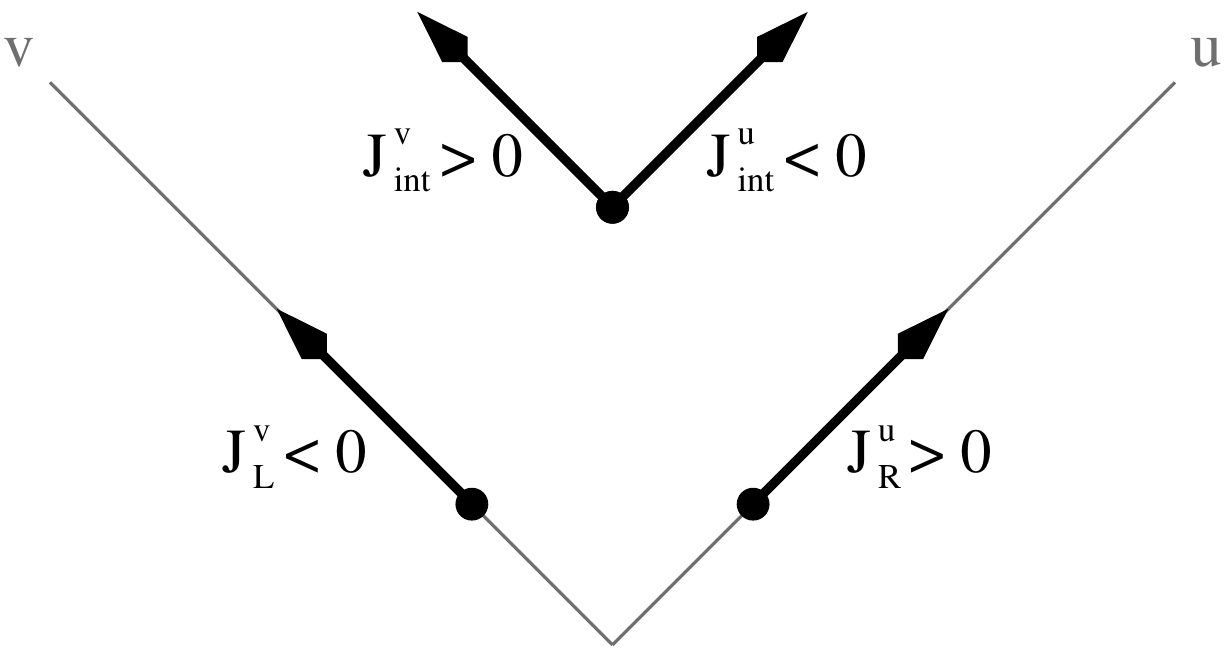}}
\caption{The quantities $J^u$ and $J^v$ represent currents flowing in
  the $u$ and $v$ directions, respectively.  Since $\Jsr$ is positive
  and $\Jsl$ is negative, both represent right-going currents. Since
  $\Jintr$ is negative and $\Jintl$ is positive, both represent
  left-going currents.}\label{fig:vectors}
\end{figure}

The behavior of the interior and singular currents is summarized in Fig.~\ref{fig:vectors}.

\section{The current contained in a strip}\label{sec:stripcur}

Since the singular current is right-going while the interior current
is left-going, it is interesting to know how much these currents cancel. For example, suppose we measure the current in $u$ flowing over a range of $v'$ from 0 to $v$. Then we find
\be\label{eqn:stripint}
\Irs(u,v)=\intov(\Jsr(u,v')+\Jintr(u,v'))dv'\,.
\ee
Using Eqs.~(\ref{eqn:Ef},\ref{eqn:Je2}), we may write
\be
\Irs(u,\infty)=\frac{Q}{2}\intoi\int_u^{\infty}(\delta(v')f(u')-\Eint(u',v')+\delta(u')f(v'))du'dv'\,.
\ee
The third term does not contribute for nonzero $u$.  We then integrate over $v'$ first.  In the first term we get $f(u')$. For the second term, Eq.~(\ref{eqn:Eintzero}) gives
\be
\intoi\Eint(u',v')dv'=f(u')\,,
\ee
and thus $\Irs(u,\infty)=0$.  Then
\be\label{eqn:stripJu}
\intov(\Jsr(u,v')+\Jintr(u,v'))dv'=-\int_v^{\infty}(\Jsr(u,v')+\Jintr(u,v'))dv'\,.
\ee
We may now solve for $\Irs$ in the strip for finite $v$. The result is
\be\label{eqn:stripfinal}
\Irs(u,v)=\frac{Q}{2}\frac{1}{1+Q\ln(uv/\delta^2)}\,.
\ee

Because this current is positive for any $(u,v)$, we conclude that the
right-going singular current is larger in magnitude than the total
right-going interior current for any strip of finite $v$. As the size
of the strip becomes very large ($v\to\infty$), the scattered
current will cancel out the singular current, so the total current in
the strip approaches zero.

While we have only examined the right-going current, the left-going
case is analogous.

\section{The total current}\label{sec:totcur}

We have found the current everywhere. It is now useful to define the
total currents $\Jrt$ and $\Jlt$ by combining our solutions. Because
these solutions are for a $\delta$-function applied field
$\Eext$, the total currents are the Green's function current
solutions. They are given by
\blea
\Jrt(u,v) &=&\Jsr(u,v)\theta(u)+\Jintr(u,v)\theta(u)\theta(v)\,,\\
\Jlt(u,v)&=&\Jsl(u,v)\theta(v)+\Jintl(u,v)\theta(u)\theta(v)\,,
\elea
where the Heaviside $\theta$ functions ensure that we only look at
points in the future of the origin.

We use this result to consider the effects of a non-singular $\Eext$. If we apply an initial electric field over a finite space-time region, the resulting charges, currents, and fields can be found integrating the solution for the $\delta$-function source. For a general externally applied field $\Eext$, the current in $u$ is given by
\be\label{eqn:genjint}
\Jr(u,v;\Eext)=\int_{-\infty}^{\infty}\int_{-\infty}^{\infty}\Eext(u',v')\Jrt(u-u',v-v')du'dv'\,,
\ee
and the current in $v$ is analogous.

\begin{figure}
\centerline{\includegraphics[scale=0.75]{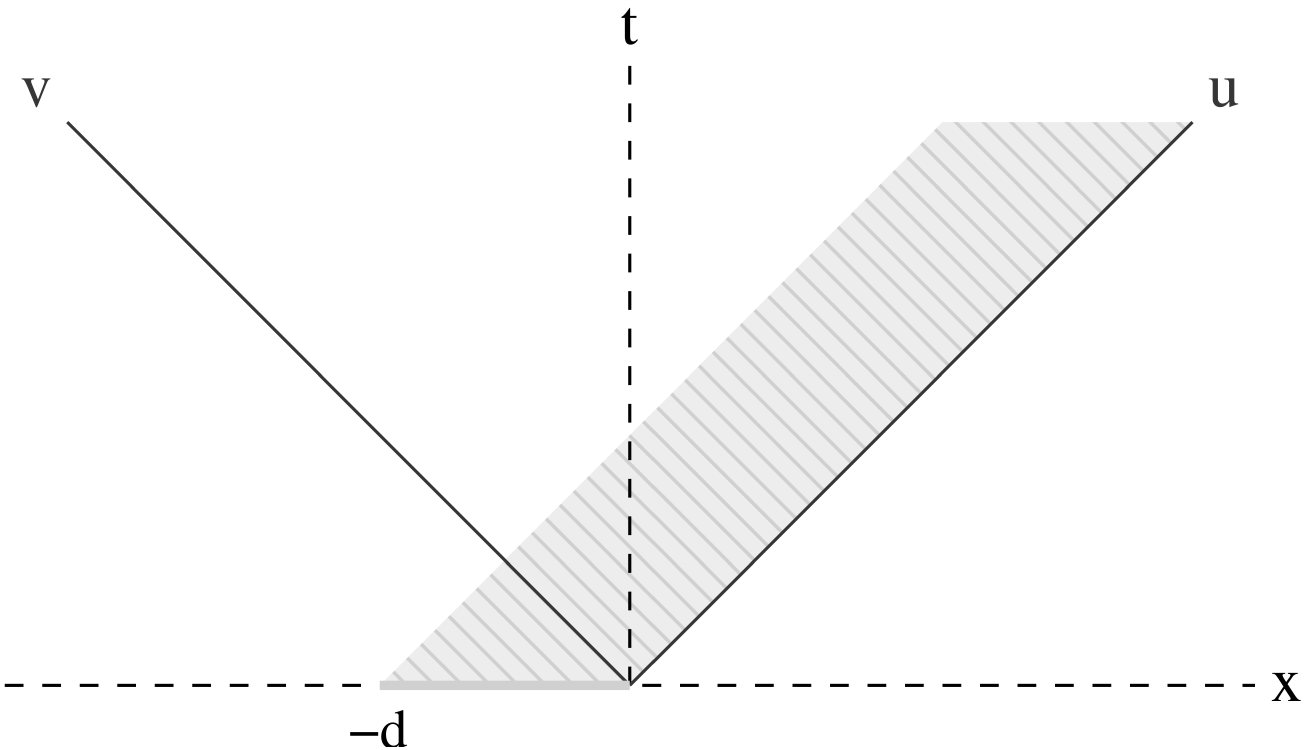}}
\caption{An electric field is applied along the $x$ axis from $-d$ to
  $0$. We are interested in the shaded region, which indicates points
  $u$-like connected to the initial distribution.}\label{fig:tophat}
\end{figure}

Now, we will consider the case where the applied electric field is a top-hat distribution with width $d$,
\be
\Eext(x,t)=\delta(t)(\theta(x+d)-\theta(x))\,,
\ee
or
\be
\Eext(u,v)=\sqrt{2}\delta(u+v)(\theta(u-v+\sqrt{2}d)-\theta(u-v))\,.
\ee
If we first integrate over $u'$, then Eq.~(\ref{eqn:genjint}) becomes
\be\label{eqn:genju1}
\Jr(u,v;\Eext)=\sqrt{2}\int_{0}^{d/\sqrt{2}}\Jrt(u+v',v-v')dv'\,.
\ee
Consider the current in a region that is $u$-like connected to the source distribution, $v\leq d/\sqrt{2}$, as shown in Fig.~\ref{fig:tophat}.
If we take $u\gg d$, then we may neglect $v'$ in comparison to $u$, and Eq.~(\ref{eqn:genju1}) becomes
\be\label{eqn:genju2}
\Jr(u,v;\Eext)=\sqrt{2}\int_{0}^{d/\sqrt{2}}\left(\Jsr(u,v-v')+\Jintr(u,v-v')\theta(v-v')\right)dv'\,.
\ee
If we then consider where $\Jsr(u,v-v')$ and $\theta(v-v')$ are supported, we may change the upper limit of integration and rewrite Eq.~(\ref{eqn:genju2}) as
\be
\Jr(u,v;\Eext)=\sqrt{2}\intov\left(\Jsr(u,v-v')+\Jintr(u,v-v')\right)dv'\,.
\ee
Using Eqs.~(\ref{eqn:stripint},\ref{eqn:stripfinal}), we find
\be
\Jr(u,v;\Eext)=\frac{Q}{\sqrt{2}}\frac{1}{1+Q\ln(uv/\delta^2)}\,.
\ee

To what degree does the current retain the top-hat shape of the
source?  The current at the leading edge is only the singular current;
the current at the trailing edge is lower by factor
\be
\frac{1+Q\ln(u/\delta)}{1+Q\ln(u/\delta)+Q\ln(v/\delta)}\,.
\ee
This is never less than $1/2$ and goes to 1 as $u$ becomes
exponentially larger than $v$.  So there is some distortion, but it is
never large and decreases as the current diminishes over time.

\section{Discussion}\label{sec:concl}

When an electric field is applied to a superconducting string, it
induces a current, but the current is reduced because of the effect of
self-inductance \cite{CSOTD}, which one can interpret as the need to
create not only the currents on the string but the associated
electromagnetic fields. The self-inductance is $L=2\ln(R/\delta)$
where $R$ is some radius cutoff. However, in the case of a straight,
static string exposed to a momentary field, there is no cutoff, so the
self-inductance is formally infinite, and no persistent current can be
produced. This is another way of describing the paradox that any
persistent current would carry with it an infinite amount of field
energy.

We found above that currents are induced initially, but they then drop
inversely with $\ln(u/\delta)$ or $\ln(v/\delta)$. The
inverse-logarithm drop-off cancels the logarithmic divergence that
would otherwise occur in the electromagnetic field energy, preventing
the paradox.

Since $\ln(u/\delta)$ increases rapidly when $u\sim\delta$, but then
more and more slowly for increasing $u$, the initial current drops off
quickly at the start but the decrease becomes less and less when
$u\gg\delta$. Half of the singular current has vanished when
$u=\delta e^{1/Q}$. The time at which a fraction $f$ of the
right-moving singular current remains is given by $u=\delta
e^{(1-f)/(Qf)}$.

\begin{figure}
\centerline{\includegraphics{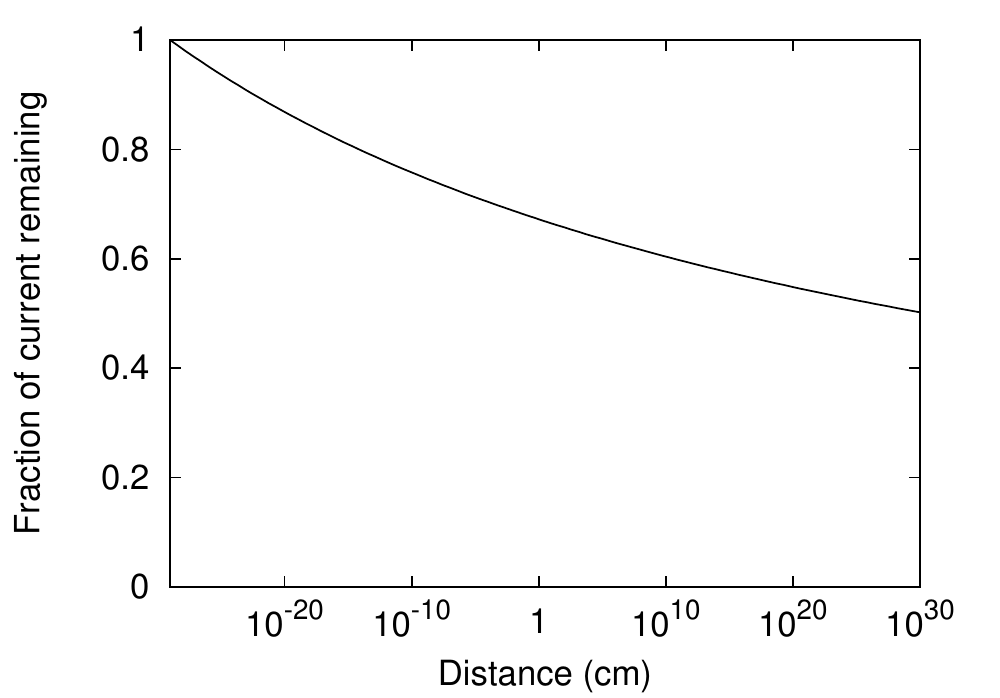}}
\caption{The fraction of current remaining falls as the inverse of the
  logarithm, so the effect is observable right away, but never becomes
  huge. We assume a string thickness of $\delta=10^{-29}\,\text{cm}$.}\label{fig:logs}
\end{figure}

Now let $\mu$ be the cosmic string tension.  Multiplying by Newton's
constant $G$ gives a dimensionless measure $G\mu$, which must be less
than $10^{-8}$ to avoid conflict with pulsar timing observations
\cite{Sanidas:2012ee,gmulim}. Strings of this scale would have a
thickness $\delta\approx10^{-29}\,\text{cm}$.  For this $\delta$, half
of the current has vanished when $u$ is about the size of the
observable universe, so the back-reaction never decreases the current
more than this. On the other hand, smaller decreases start quite
rapidly; one quarter of the current has vanished when $u$ is about one
angstrom. Figure~\ref{fig:logs} shows the rapid initial drop-off of
the current, followed by the long tail.

This study of electromagnetic back-reaction may be of theoretical
interest in examining the gravitational back-reaction of a string. In
this scenario, a static straight string is suddenly displaced in space
in a local region. We can then predict how the left- and right-going
kinks generated by this displacement evolve over time. We expect the
effect of gravitational back-reaction to be similar to what we found
here for electromagnetism, except that in place of the squared charge
$Q$ the gravitational self-coupling will be proportional to
$G\mu$. For a string with $G\mu\approx10^{-8}$, the kinks would be
damped by about one part in one million at the length scale of the
observable universe. Thus, we would not expect any observable damping
due to this effect in the gravitational case.

These results are applicable to straight strings and concern the
divergent effects associated with a string of zero width or an
infinite transverse space.  In the case of loops, or infinite strings
that are not straight, there will be other effects, which are not
divergent \cite{Quashnock:1990wv}, but nevertheless are likely to be
larger in realistic situations than the effects we discuss here.

\section*{Acknowledgments}

We are grateful to Alex Vilenkin, who suggested solving the simpler
electromagnetic case before looking at gravitational
back-reaction. This work was funded in part by the National Science
Foundation under grants number 1213888 and 1213930.

\appendix
\section{Analytic structure of $F$ and $\mathcal{H}$}

In order to take inverse Laplace transforms, we need to know the
analytic structure of the functions $F$ and $\mathcal{H}$.  First
consider $F$, given by Eq.~(\ref{eqn:F}).  The function
$\Eone(z)$ has a logarithmic singularity at $z=0$ and a branch cut on
the negative real axis, and is otherwise analytic.  In
Eq.~(\ref{eqn:F}) the logarithmic singularity cancels between the
numerator and the denominator, so there is no divergence at the
origin, but the branch cut remains.

The only possibility for additional singularities in $F$ not present
in $H$ is for the denominator of Eq.~(\ref{eqn:F}) to vanish.  We can
show that this cannot happen as follows.  From the definition of
$\Eone$, we can write $e^z\Eone(z)=\int_z^\infty dt\, e^{z-t}/t$ and
take the contour to run first in the positive real direction and then
in the imaginary direction at positive real infinity, where it does
not contribute.  Thus $\Eone(z)e^z=\intoi dx\, e^{-x}/(z+x)$.  For
positive real $z$, the result is positive.  For $z$ with positive
imaginary part, $1/(z+x)$ always has negative imaginary part, and vice
versa, so nowhere in the domain of definition of $\Eone$ can
$e^z\Eone(z)$ be a negative real number.  Thus $1+H(z)$ can never
vanish.

Deformation of the integration contour from $-i\infty\ldots i\infty$
to enclose only the negative real axis yields circular contours at
radius $r\to\infty$.  As $w\to\infty$, $\Eone(w)\approx e^{-w}/w$.
Thus, $F(w)\sim1/w$ for large $|w|$.  Thus a contour of radius $r$
yields $\int_0^{\pi/2} r d\theta \exp(-uw)/w$.  The magnitude of the
integrand is less than $\exp(-ru\theta)$, so there is no contribution
as $r\to\infty$.

Now we turn to $\mathcal{H}$, given by Eq.~(\ref{eqn:HH}).  The novel
feature here is that there can be a singularity if $H(w)
+H(z) = -1$, which can indeed happen (though only if $Q$ is
larger than 0.67, because the minimum value of $\Real e^{-z}\Eone(z)$
is about $-0.74$).  Nevertheless, it is still possible to deform the
contours in Eq.~(\ref{eqn:L2}).  For each $w$, let us deform the
contour in $y$.  We may discover some isolated points $y_i$ where
$H(y_i) = -1-H(w)$.  At such a point, there will
be contributions of the form $2\pi i/H'(y_i)$.  Now we will
deform the contour in $w$.  If there is any point where
$H'(y_i) = 0$, there will be a contribution from deforming
the contour across a pole.  But in fact there is no such point.  We
have
\be
H'(y) \propto e^y\Eone(y) - \frac{1}{y}
= \intoi dx\, e^{-x}\left[\frac{1}{y+x}-\frac{1}{y}\right]
= -\frac{1}{y}\intoi dx\, e^{-x}\frac{x}{y+x}
\ee
If $y$ is positive and real, the integrand is always positive.  If $y$
has positive imaginary part, then the integrand has negative imaginary
part, and vice versa.  So the integral can never vanish.  Thus there
are no poles in $1/H'(y_i)$.

As $y\to\infty$ with $w$ fixed, $\mathcal{H}\sim 1/y$, so there is no
contribution from contours at infinity, as above.  Once we have deformed
the $y$ contour, we can deform the $w$ contour with again a
contribution $\sim 1/w$ in the $w\to\infty$ limit.  Thus there is no
obstacle to reaching Eq.~(\ref{eqn:deform2})

\bibliography{paper}

\end{document}